# Broadcasters and Hidden Influentials in Online Protest Diffusion

Sandra González-Bailón, Javier Borge-Holthoefer and Yamir Moreno

University of Oxford and University of Zaragoza

Authors Note

Sandra González-Bailón, Oxford Internet Institute, University of Oxford

Javier Borge-Holthoefer and Yamir Moreno, BIFI, University of Zaragoza

This research was supported by the Spanish MICINN projects CSO2009-09890, CSD2010-00034, and FIS2008-01240 and FIS2009-13364-C02-01, and by the Government of Aragon (DGA) through the grant No. PI038/08. We thank Cristina Viñas and Alejandro Rivero for their help in preparing the data.

Correspondence concerning this article should be addressed to

Dr. Sandra Gonzalez-Bailon, Oxford Internet Institute, University of Oxford, 1 St. Giles, Oxford, OX1 3JS, telephone: +44 (0) 1865 287 233, fax: +44 (0) 1865 287 211, e-mail: sandra.gonzalezbailon@oii.ox.ac.uk.




## Abstract

This paper explores the growth of online mobilizations using data from the 'indignados' (the 'outraged') movement in Spain, which emerged under the influence of the revolution in Egypt and as a precursor to the global Occupy mobilizations. The data tracks Twitter activity around the protests that took place in May 2011, which led to the formation of camp sites in dozens of cities all over the country and massive daily demonstrations during the week prior to the elections of May 22. We reconstruct the network of tens of thousands of users, and monitor their message activity for a month (25 April 2011 to 25 May 2011). Using both the structure of the network and levels of activity in message exchange, we identify four types of users and we analyze their role in the growth of the protest. Drawing from theories of online collective action and research on information diffusion in networks the paper centers on the following questions: How does protest information spread in online networks? How do different actors contribute to that diffusion? How do mainstream media interact with new media? Do they help amplify protest messages? And what is the role of less popular but far more frequent users in the growth of online mobilizations? This paper aims to inform the theoretical debate on whether digital technologies are changing the logic of collective action, and provide evidence of how new media facilitates the coordination of offline mobilizations.

## Introduction

The year 2011 was punctuated by the emergence of protests in several countries around the world. The January uprisings in Tunisia were soon followed by social unrest in many other countries of the MENA region, leading to the revolutions of Egypt and Libya and shaking the foundations of several other dictatorial regimes, a set of uprisings that has come to be known as the Arab Spring. This wave of dissent in authoritarian states soon extended to liberal democracies, with citizens in Spain, Greece and Chile staging massive demonstrations against their political leaders, Israel protesters advocating for greater social justice, and US demonstrators setting camp sites in the squares of several cities, following the original occupation of Zuccotti Park in New York. These mobilizations paved the way for the global Occupy movement that brought, towards the end of the year, thousands of protesters to the streets of hundreds of cities worldwide under the slogan "We are the 99%". In the process, protesters consolidated new tactics in the social movement repertoire, like camping in public spaces and creating on-site media centers to use online networks for information diffusion.

The prominence of these events was epitomized by the decision of *Time* magazine to dedicate their 2011 Person of the Year issue to the protester. According to the editors, the word 'protest' appeared in newspapers and online "exponentially more in 2011 than at any other time in history". Because the leadership of these movements came from the bottom-up, not from the top of an organization, the editors chose the anonymous protester rather than a particular individual, highlighting the role that technology played as a crucial aide in the mobilizations: internet-enabled forms of communication, claims the report, allowed people to watch what was happening in real time and helped spread "the virus of protest" (Stengel, 2011). Although online networks did not cause the movements, the report states, they kept them alive and connected.



This decision to nominate the anonymous protester as Person of the Year closes a cycle of news reporting where online technologies have been repeatedly identified as the backbone of the protests. Most news reports highlight the prominent role of social media in the emergence and the coordination of offline mobilizations, consolidating as conventional wisdom the idea that new media is inherently linked to social unrest and popularizing expressions like "Facebook Revolution" and "Twitter Revolution" as shorthand for the popular uprisings. However, there are many open questions about how these and other online networks facilitate the emergence and diffusion of protests: How does information spread in online networks? How fast? Do online networks really promote a decentralized diffusion of information? Who are the most influential users? Journalistic accounts of the protests provide valuable insights into the personal stories of the protagonists and the narratives of the events they contributed to trigger. But understanding how social media drives the emergence of collective action on a large scale requires going beyond the contingencies of each case to the general principles that drive those dynamics.

This paper offers that type of analysis, providing an empirical examination of the diffusion mechanisms that drove online activity in one instance of mass mobilization: the protests that took place in Spain in May of 2011. The study draws theoretical insights from previous research on networks and collective action, paying special attention to how new media has changed the costs of mobilization and coordination. The focus of our analysis lies on the diffusion of protest activity through an online network, which in this case has a clear correspondence with massive offline mobilizations: the Spanish 'indignados' movement offers a particularly good example of how diffusion online is often accompanied (and can even drive) offline diffusion. The analyses that follow aim to indentify core actors driving the spread of information, exploring the relationship between new media and old media, and the influence



that prominent but also common users had in the diffusion process. Our argument is that the growth of digitally-born protests depends on the strategic deployment of preexisting networks, and on the ability to capitalize on the visibility of the best connected actors. The mechanics of this process are consistent with more generic principles of diffusion and collective action; the peculiarity of online protests is that they can scale-up quicker (by exposing people to information they might not have encountered otherwise) and they can adapt their framing to changing circumstances in a more responsive manner. The following section elaborates on the generic principles of collective action in networks to then go into the details of the Spanish case.

## Networks, Diffusion, and Collective Action

One of the key questions that have puzzled researchers of collective action for decades is what makes people contribute to public goods when, by virtue of being public, they can be enjoyed without having to contribute to their provision. The problem, as conceptualised in Olson's classic approach (Olson, 1965), arises from the assumption that individuals are rational actors motivated by self-interest. Pondering the costs and benefits of participation, rational actors find strong incentives to free-ride and end up falling in the trap of a social dilemma: they know they would be better off if the public good were produced, but they would rather have others making the effort to actually produce it; if everybody reasons along these lines, no public good is provided and everybody is worse off. This social dilemma has received attention from a number of angles: in policy research, for instance, it adopts the form of the Tragedy of the Commons, where short-term individual interests are myopically prioritised over the long-term group interests, to the detriment of all (Hardin, 1982; Ostrom, 1990); in electoral research, it is expressed in the form of the voting paradox (Downs, 1957): why do people vote when the impact of their ballot is negligible in the final outcome, and therefore, the costs of the action



exceed the benefits? In the case of social movements, the question is what makes people

participate in protests and take part in the organisation of collective demands when they could

enjoy the benefits without having to sustain the efforts.

And yet fee-riding does not always become the dominant strategy, something the recent

wave of mobilizations illustrates quite well. Most research considering the question of why

people engage in collective action in spite of their rational interests assumes a deviation from

rationality and its predictions, like acting under the effects of norms, group pressure or social

influence (Coleman, 1990; Elster, 1989). Most of this research also assumes that actors are not

isolated decision-makers but are instead embedded in densely knitted networks of social

interactions that allow the efficient enforcement of norms and the spread of reputation. This

assumption agrees with one of Olson's main claims: that unless groups remain small or they

devise the mechanisms to make individuals act in the common interest (by means of, for

instance, selective incentives) rational and self-interested individuals will still be tempted to

defect and free-ride.

Networks, however, matter not only because they allow self-regulation and control, but

also because they boost a sense of efficacy that motivates individual contributions. The

question that actors pose themselves when confronted with the decision to join a collective

effort, the argument goes, is not so much whether it is beneficial but rather if it is efficient, a

question that depends on how many other actors have already joined (Gould, 1993; Macy,

1991; Marwell & Prahl, 1988; Oliver & Marwell, 1988; Oliver, Marwell, & Teixeira, 1985). In

most empirical settings, actors do not decide in parallel but sequentially; this allows them to

see how many others are contributing before making a decision of whether to contribute as

well. Since actors are heterogeneous in their inclination to participate (they have different

thresholds with respect to how many others need to be participating before deciding to join the



action, see Granovetter, 1978 and Valente, 1996), sequential decisions allow actors that did not consider joining in a given time to join later, once a critical mass has been reached. Social influence and interdependent decision-making activate chain reactions that reduce the need for selective incentives: under the effects of social influence, collective action becomes more a process of contagion than of incentive design.

This contagious dimension makes collective action similar to other processes of diffusion (Young, 2003, 2009). Networks are crucial to understand the transmission of behaviour because they define the group of reference that individuals monitor prior to making a decision, and therefore provide the infrastructure on which diffusion takes place. Two actors with the same threshold but different personal networks might join the collective effort at different times: all else equal, actors with larger networks will require more time to register a critical mass (Valente, 1996; Watts & Dodds, 2010). Because of this, networks not only provide a structure of interdependence; some of their features, like size, density or centralisation, also affect the speed and reach of chain reactions (Gould, 1993; Marwell & Prahl, 1988; Oliver & Marwell, 1988; Siegel, 2009). A handful of highly motivated actors are necessary to start the chains, but their position in the network, and the position of those they are connected to in the network, are also relevant for the way those chains unfold.

Diffusion models help see the provision of public goods not as a binary event where the good is either produced or not, but as a more continuous process where what matters is the proportion of users contributing, and the time it takes to reach a critical mass of participants. The interdependence of individual decisions and the cascading effects they generate go beyond the parameters of rational calculation, adding a complexity to the emerging dynamics that cannot be captured just in terms of individual costs and benefits. This focus on diffusion relaxes rationality demands of classic approaches by assuming that actors learn through



experience, "adapting their decisions in response to social feedback" (Macy 1991: 731); and it is consistent with the importance of social norms: individual thresholds often respond to normative principles like fairness, for instance, as when actors are willing to contribute only in proportion to what others are contributing (Gould 1993: 183). Participation, therefore, is contingent to the actions of others and to when and how much they are willing to contribute. Although networks do not solve the initial "volunteer's dilemma" – i.e. who decides to take part first – they make participation spread; it is this sort of social influence what sustains the decentralised nature of many protests and mobilizations.

Empirical examples of these network dynamics have been found in the context of insurgencies, political demonstrations, the growth of unions, contentious action, and voting behaviour (Biggs, 2005; Gould, 1991; Hedström, 1994; Lohmann, 1994; Rolfe, 2010). The wave of protests that took place in 2011 provides an excellent empirical ground to assess whether these dynamics are really changing in the digital era. The one common denominator of these protests (which, in all other respects, differ widely in the contingencies imposed by their local contexts) is that they emerged without the structure of formal organizations, involving large numbers of people that were recruited and mobilized using online networks. This challenges two key assumptions of traditional theories of collective action: one, that the costs of participation encourage free-riding behavior; and two, that formal organizations and small groups are needed to encourage individual contributions. The new communication environment created by internet technologies has significantly reduced the costs of participation, to the point that they make the free-rider problem less problematic and the role of organizations in enforcing sanctions and selective incentives less meaningful (Bimber, Flanagin, & Sthol, 2005; Lupia & Sin, 2003). Online networks allow communities to self-organize without the need of formal structures or co-presence, and they create the means for the diffusion of novel tactics that widen the protest repertoire (Earl, 2010; Earl & Kimport, 2011; Shirky, 2008). However,



the diffusion patterns followed by these protests also fall in line with the larger body of

research on diffusion in networks, suggesting that the same mechanisms of social influence,

learning or contagion might be at play. Focusing on this dimension can uncover more

similarities between online and offline collective action than has been acknowledged so far.

The two main aspects of diffusion – the activation of thresholds at the individual level,

and the chain reactions triggered at the collective level – operate online as well as offline. More

central actors or actors with similar network positions can be more consequential for diffusion

processes because of the distribution channels that their local networks grant them (Burt, 1987;

Iyengar, Van den Bulte, & Valente, 2011; Marwell & Prahl, 1988; Valente, 1996). Resource

mobilization theories have actually highlighted the importance of network centrality to reach

and mobilize resources that are essential for the success of social movements (Diani &

McAdam, 2003). Pecuniary costs are not as relevant online, but scarcity of attention and the

size of audiences still matter, particularly when it comes to exposing people to information

they are unlikely to encounter otherwise. For this, some actors have more chances of success

than others because of the networks in which they are embedded.

Another crucial question in the analysis of diffusion dynamics is how common

exposure to a global source of information (usually in the form of traditional media, or

marketing campaigns) interacts with the transmission of information in local networks.

Previous studies have shown that what seemed to be an example of diffusion (Coleman,

Menzel, & Katz, 1957) was in fact the result of a marketing campaign (van den Bulte & Lilien,

2001), that is, a source of pressure exogenous to the network. The common assumption in most

accounts of the 2011 protests is that they were driven by online communication; however, big

media outlets like AlJazeera, the BBC or the CNN, were also covering the events. There is

evidence of past riots and protest waves that underscore the importance of mass media in



driving the diffusion of the protests (Myers, 2000). Disentangling the role of new media in social unrest requires taking into account the way traditional media interacted with the events, both as offline broadcasters and as online actors.

Diffusion models allow measuring the success of collective action in relative rather than absolute terms, that is, as the fraction of a given population that embarked in a given course of action, and the time it took to get there, given network effects and externalities. Focusing on the diffusion dynamics opens a more useful approach to make sense of the sort of mobilizations that emerge online, which are more transient than those traditionally supported by established organizations (Bennett, 2003), and have a less unified purpose, as instances of flash activism illustrate (Earl & Kimport, 2011: 73). Diffusion models give us a set of tools to understand the dynamics of group formation and growth, and the mechanisms through which a movement reaches new subsets of the population. Instead of resources and selective incentives (which are the two theoretical building blocks emphasized by classic approaches to collective action), diffusion models base their approach on social influence and network dynamics; consequently, they offer a more natural frame to understand collective action in the digital era, when the boundaries between private and public are more porous and personal networks can be easily used for public purpose (Bimber 2005: 378). Implicit in most recent approaches to digitally-enabled collective action is the assumption that the internet is changing the costs-benefit calculation of potential actors. Here we focus on the less explored dimension of how interdependent decisions trigger chain reactions that end up building a critical mass of participants.



## The Spanish Case in the 2011 Wave of Protests

The Spanish 'indignados' ( 'outraged') movement is a step in the sequence of events that went from the Arab Spring in the MENA region at the beginning of 2011 to the Global Occupy movement towards the end of the year. The movement emerged as a civic initiative with no party or union affiliation that protested against political alienation and demanded better channels for democratic participation. The first big demonstration took place on May 15, and it was organized by the digitally coordinated platform Democracia Real Ya ("Real Democracy Now"), born online about three months before that first demonstration day. Hundreds of entities joined the platform, from small local associations to territorial delegations of larger groups like ATTAC (an international anti-globalization organization) or Ecologists in Action. Signatories of the original call included student associations, bloggers, defenders of human rights and people from the arts, but also hundreds of individual citizens of different age and ideologies. The motto of the movement was "take the streets", with other slogans including "we are not goods in the hands of politicians and bankers" or "we don't pay this crisis".

The protests of May 15 brought tens of thousands of people to the streets of more than fifty cities all over the country. Figure 1 gives a snapshot of the online (Twitter) activity generated by users in these cities. After the march, some demonstrators decided to continue the protests by camping on the squares of the main cities until the following Sunday, May 22, the date for regional and local elections. During that week, the authorities tried to evict camped protesters by force, and the Electoral Committee declared the protests illegal, but these events only increased the media visibility of the movement and boosted popular support. After the elections, the movement remained active, organizing another big demonstration later in the year, on October 15, under the motto "United for Global Change", this time part of the global Occupy movement.



-- Figure 1 about here --

These protests were greatly inspired by the Arab Spring. In the words of one of the protesters, "After seeing what happened in the Arab countries, you ask yourself, why not here? If they can do it there, where things are so much worse, don't we, with our supposed democracy, have a responsibility to try to make things better as well?" (Andersen, 2011). The Spanish protests inspired, in turn, subsequent mobilizations in Greece, Chile, and the Occupy movement of New York and soon replicated in other cities. In the process, strategies and tactics were exchanged through online networks (Ibid.). This suggests that there was a spatial diffusion driven by overlapping channels of communication that spanned geographical borders. However, within each country there were also high volumes of online activity that allowed each of these instances of mobilization to brew and ultimately explode. By focusing on one of these local networks, the Spanish movement, we aim to open an avenue for research that can ultimately be expanded to understand the global patterns of diffusion of which the Spanish experience offers just one case.

The analyses that follow aim to answer the following questions: How does protest information spread in online networks? How do different actors contribute to that diffusion? Do traditional media lead or follow information flows? How do they interact with new media? And what is the role of less popular but more frequent users in the growth of the mobilization? In the light of the discussion presented in the previous section, there are two aspects that are important to understand the growth of this movement: the structure of the communication network (and the position that different actors occupy within that structure), and the chain reactions triggered by the flow of protest messages (which create bursts of activity in the network). In line with previous models of network diffusion, we should expect more central



actors to be more consequential in the dissemination of protest information. However, if recent accounts of online mobilizations are correct, the coordinating power of online networks would actually derive from their decentralized structure, and the possibility they grant to circumvent traditional media hubs. The following section presents the data used to determine which of these two possibilities is best supported by the evidence of this case.

## Data and Methods

Our data consists on Twitter activity around the protests for the period April 25 to May 25. This observation window goes back a few weeks before the first big mass demonstrations, which allows tracking online activity before the movement became visible in mainstream media. The method to monitor Twitter activity around the protests was applied in two stages. First, we selected hashtags that were relevant to the protests, coming up with a list of 70 keywords. Figure 2 shows the most prominent tags in terms of frequency of use prior and after the demonstration of May 15. The evolving salience of these tags gives some insights into how the movement framed itself during its emergence and growth. Before 15-M, most protest messages are tagged with a reference to the demonstration ("15m", "tomalacalle" or *take the streets*), the online platform promoting it ("democraciarealya" or *real democracy now*), and the main message of the protest, namely, the demand for new forms of democratic representation ("nolesvotes" or *don't vote for them*). On May 15, the day of the first mass demonstrations, other hashtags gained prominence: "spanishrevolution", "acampadasol", "acampadabcn", "acampadasevilla" and "acampadavalencia" (the last four as an explicit reference to the camps set up in Madrid, Barcelona, Seville, and Valencia, respectively). Other tags ("nonosvamos" or *we don't leave* and "yeswecamp") emerged later as a response to the attempts to evict the squares. Towards the end of the observation window, when the elections have already taken



place, new hashtags (i.e. "consensodemininos" or *minimum consensus*) signal the evolution of the movement into a new, more deliberative stage.

We collected messages that included these hashtags, with the constraints that we only archived messages written in Spanish and sent from Spanish territory. We estimate that our sample captures about a third of the total number of tweets related to the protests, amounting to a total of more than half a million messages. This sample of tweets contains a subset of messages that were targeted at other users, that is, messages that in addition to a protest hashtag also used the @ symbol to identify other people by their username. This subset of messages capture more directly the communication flows between users and is central in the analyses that follow.

-- Figure 2 about here --

The second stage of data collection involved reconstructing the following/follower network of the users sending protest messages. We used their ids (i.e. unique identifiers) as the starting point of a crawl that applied a one-step snowball sampling procedure and identified all neighbors directly connected as users following or being followed. This data allows working with two versions of the network: one is the original follower network, where some connections are not reciprocated and open only one-way channels of communication; the second is a symmetric version that only retains mutual arcs and therefore two-way communication channels between users. Distinguishing these two networks allows us to identify users we refer to as "hubs", that is, very prominent users that have a long list of followers (and therefore large networks) but who only reciprocate a small fraction of those connections. These hubs are typically public figures or celebrities, and they are less prominent



in the symmetric network. The following analyses aim to uncover the relevance of these hubs in the diffusion of protest information.

These data track a substantive volume of online activity around the protests, but they only allow us to account for one part of the story: this mobilization had a very strong presence in the streets and high visibility in mainstream media, particularly after protesters decided to set up camps prior to the Election Day. Since we are only analyzing what happened in an online network we are missing many channels for diffusion, for instance those opened by offline networks or exposure to mass media. And yet, to the extent that this organization (Real Democracy Now) was born online, our data allow us to analyze the origins of the movement: it ultimately exploded to become prominent offline but it would have never emerged with such speed and scale in the absence of online communication networks. This is supported by figure 3, which  displays the chronological growth of the movement as measured by number of active users and protest messages, and compared to the number of headlines in traditional newspapers mentioning the protests (according to the database Nexis and Google news).

-- Figure 3 about here --

Figure 3 shows that the movement was virtually invisible in mainstream media until it had already exploded in the form of mass demonstrations. This lack of media coverage means that, until the movement took to the streets, most information was disseminated online. The effects of offline communication networks, as mediated by news media, are practically negligible until at least 20% of users had already joined the online exchange of protest information. This suggests that online communication networks are now playing the role that news media played in past protests and waves of contention (Myers, 2000). The figure also shows that the movement left a digital footprint typical of diffusion processes (Young 2009):



an S-shaped curve with an initial phase of slow growth, a phase of quick escalation, and a

burnout phase, when most users had already sent at least one protest message and the diffusion

approaches saturation in this population of users.

**Analyses**

The platform behind these protests was born online as a virtual assembly of otherwise

dispersed actors and organizations. Our question is how their message was diffused to the

larger population, and whether and how online networks helped in that diffusion. We pay

attention to both the structure of the Twitter network and to the dynamics of message exchange

this network facilitated. The scatterplot in Figure 4 summarizes how users distribute in the

network of followers and in the allocation of targeted messages. Both axes are expressed as

ratios so that it is easier to identify outliers, that is, users who depart from symmetrical

networks or from the volume of message exchange entailed by mere reciprocation. The vertical

axis tracks the number of messages that users received over the number of messages they sent:

the most visible users (those who were mentioned more often in protest messages) are above

the dashed line. The horizontal axis tracks the number of other accounts a user is following

over the number of followers they have: the most central and popular users in this

communication network are on the left of the dashed line.

-- Figure 4 about here --

The color of bins is proportional to the number of users that fall in that area. What the

scatterplot shows is that most users active in these protests are average in terms of their levels

of activity and the size of their networks: they receive roughly the same number of messages

that they send, and they have roughly the same number of followers that they follow (although



their networks tend to be asymmetrical in favor of hubs or celebrities). These average users are identified around the intersection of the dashed lines. The second thing the scatterplot shows is that hubs receive more targeted messages than normal users: most of the activity goes towards celebrity accounts, identified in quadrant 1. We call these users "influentials" (N=4048) because they are central both in the overall communication network and in the domain-specific communication exchange of protest messages: other users direct their messages to them in the hope that they will pass them on and help them reach a larger number of people.

The figure also displays a less obvious finding: there is a portion of users who, in spite of being average in their network centrality, still attract a high number of protest messages. They are located in quadrant number 2, and we call them "hidden influentials" (N=8472) because there is nothing about their local networks that would a priori identify them as particularly visible; and yet they are very visible in the context of these protests. Users in quadrant number 3 (N=3309) are general broadcasters who send more messages than they receive and have the potential to spread them more widely given their higher number of followers. And, finally, in quadrant 4 we have the largest set of common users (N=30173) who send more messages than they get and have a relatively small audience.

Table 1 sheds light into who are the most prominent actors in each of these categories: it contains the list of top 20 users both in terms of their network centrality and in terms of the number of protest messages that they receive, overall and for each of the four subsets identified in Figure 4. As expected, users in the top ranks of network centrality are celebrities or established news organizations; there are also prominent users (in terms of number of followers) that are irrelevant for the protests, like a sports club and two procrastinating accounts, which indeed disappear from message visibility rankings. Celebrities and news organizations are again the most prominent users in terms of global message visibility.



However, this ranking is also populated by a new media site (YouTube) and Twitter applications that allow posting videos, pictures and facilitate the diffusion of information. Only one account created as part of the protests, *acampadasol*, made it to this global top 20: it refers to the first camp set up by protesters in the main square of Madrid, the epicenter of the movement.

-- Table 1 about here --

The top users in terms of global message visibility are also very central in the overall Twitter network, this is why they appear again as the top 20 "influentials". It is in the "hidden influentials" column where most of the accounts created to promote the protests appear: they are the nodes in the network that were created to embody the movement and act as the gravity centers of the protests. These accounts make an explicit reference to the first protest day (15m), the online platform behind the protest (Democra_Real_Ya or *Real Democracy Now*), and the main message conveyed by protesters (nonosvamos or *we don't leave* and indignaos or *feel outraged*). Also significantly, under this category of "hidden influentials" we find various personal accounts, that is, normal users that became prominent targets in the flux of protest information. Personal accounts are also the most visible "broadcasters", with a few celebrities, a news organization and a new media grassroots group. As the name implies, the vast majority of visible users in the category "common users" are personal accounts.

This table confirms the importance of differentiating between users who have broadcasting potential (because they are connected to a larger number of followers) and those who actually act as such in the context of specific information diffusion events, like this protest. Traditional media organizations or celebrities might be very central in terms of their network sizes, but when it comes to particular streams of information, message activity is a



more relevant measure of visibility. The broadcasters we identify here are not only more central in the network (and therefore have relatively larger audiences); they are also more active senders than receptors. Their influence, however, is not attributed by other users as in the case of the "influentials" subset: broadcasters are not as often targeted as message recipients and therefore lack their visibility. And yet they may play a crucial role as transmission belts, enhancing the visibility of those influentials we identified as "hidden" – and therefore propelling the movement to its growth.

    This classification of users leads to two questions: Who started the campaign? And who was more successful at triggering information cascades that led to the growth of the movement? Figure 5, panel A shows that there are no significant differences in activation times across types of users: most users started sending protest messages around the day of the first big demonstration. Broadcasters and common users, however, seem to have slightly higher levels of protest activity prior to that day: compared to them, influentials and hidden influentials have higher activation rates right after the first big demonstrations. This slight advantage as leaders of the movement is particularly relevant in the case of broadcasters because they had the connections that the new born "hidden influentials" lacked and helped disseminate their existence by means of targeted messages.

    Panel B in figure 5 shows the size distribution of the information cascades initiated by users classified in each of the four groups. The idea behind our operationalization of cascades is that users are more likely to send protest messages if they see any of their network neighbors doing so. This form of influence activates a chain reaction that might unfold over the network and ultimately reach a high number of people. The chain might also die in the early stages if only a few followers, or none of them, decide to send a protest message of their own after being exposed to the first message. The question of interest to understand what actors were



more relevant in the growth of the movement is who was more likely to activate long chains and therefore large cascades. What figure 5 shows is that, on average, influential users triggered the largest cascades, followed by broadcasters. Their better network connectivity (i.e. their higher number of followers) translates into a higher outreach capacity. As Table 1 showed, these users are mostly celebrities and traditional news media, but the set also includes personal accounts and, most significantly, accounts created by the protesters: *acampadasol* (which refers to the main camp set up in Madrid) is in the top 20 list of influentials. Common users and hidden influentials occasionally trigger cascades that are comparable in size (these are the outliers depicted in the upper region of the boxplots), but their ability to disseminate wide calls is notably lower.

-- Figure 5 about here --

These differences in cascade sizes explain why influentials are such a prominent target for message exchange: it makes strategic sense to try to reach them because they can make the message reach a large number of people and potentially motivate them to join the movement as well. What this means is that, in line with resource mobilization theories, protesters also try to mobilize resources in online networks, where resources take the form of access to a wider audience. While hidden influentials give the movement identity and framing (hence their visibility in the exchange of protest messages), influentials help catapult their message. In other words, protesters employ online networks both to frame the movement and to maximize outreach; our analyses show that the fulfillment of these two goals relies on a division of labor, voluntary (as when broadcasters build the visibility of hidden influentials) or induced (as when influentials are systematically targeted in the hope that they will pass the message along).



## Discussion

Political action and the provision of public goods have long puzzled social scientists because of the assumption of selfish rationality and the prevalence this assumption gives to costs in decision making. This paper has argued that reducing costs is not the only way in which the internet is changing the logic of collective action; it is also strengthening the interdependence of decision making, that is, the fact that actors decide in the context of a group of reference whose behavior influences their final decision. This makes collective action not so much a problem of providing incentives that will compensate for the costs of participation, but rather a matter of having access to information that can help assess the efficiency of participation. Communication networks channel that information and allow actors to infer how many other people are already participating; this is important not only to boost their sense of political efficacy, but also to instill in them normative behavior: participants are more likely to conform to average contributions if they know how much other people are contributing. The network dynamics of social influence further facilitate coordination by triggering information cascades that can potentially reach a high number of people in a short time span. None of these processes can unfold as efficiently offline as it does in online networks, and that is the reason why they have played such a fundamental role in the organization of recent protests.

However, this still begs the question of what are the elements in those networks that make them be so efficient. What our findings suggest is that common assumptions of horizontal structures are actually unfounded: the reason why online networks are so efficient is because they have a very centralized structure that allows reaching a high number of people with short chains of diffusion. Our analyses reveal that while the sparks that set the movement in motion are distributed all over the network (illustrated by the activation times in figure 5, panel A), the information cascades that maximize the number of people exposed to protest



information start at the centre of the network (panel B). The reach of the movement is in the hands of a few influentials and broadcasters. Hidden influentials, however, are crucial in the process because they introduce the memes, or ideas, to be distributed.

Our data also shows that online networks supplant mainstream media in the early stages of the movement: they offer the main channels for information diffusion, and allow the protests to take off. However, once the protests occupied the streets, and traditional media started to cover the events, online communication interacts with offline exposure to information, creating feedback effects that we cannot capture with our data but surely contributes to inflate the number of users that became activated (particularly during the phase of exponential growth in figure 3). However, traditional media affected the process not only exogenously but as a prominent member of the online network as well: most influentials belong to their category. And yet it is very significant that a node in this network with only a few months of existence (the *acampadasol* account, which refers to the organizational epicenter of the movement) is able to compete with prominent media outlets in terms of centrality and visibility. Much as traditional media still monopolizes most of the information flow, the fact that an account created by protesters gets so prominent in a few days provides evidence of the extent to which online networks can help break the logic of preferential attachment, that is, the tendency to reinforce the centrality of already central users. Exceptions like this are important in the gestation of exceptional events: the movement managed to get to the core of a network highly centralized around celebrities and traditional media hubs.

Overall, this study suggests that we need to conceptualize collective action less as the dynamics that emerge from binary decision making around costs-benefits calculations, and more like a diffusion process where what matters is how many others are already in, and how much exposure they have to undecided actors. To the extent that networks define that exposure,



they offer the key to understand a crucial mechanism in the emergence of collective action, namely the activation of individual thresholds, which in turn can trigger chain reactions on a global scale. Online networks are not unique in facilitating these processes – offline communication networks are very good at channeling influence as well – but they are certainly more efficient. This, however, does not necessarily mean that they can always fulfill their function in facilitating collective action: online networks offer, in the end, only one of the many layers of the structures in which people are embedded. This in turn means that online networks or, new media more generally, do not hold the key to predict the next wave of protests. In hindsight, though, we can trace back how networks helped a movement grow out of nowhere, and this can definitely help shed light into why, in spite of the reservations of some pundits (Gladwell, 2010), the revolution was indeed tweeted only a few months down the line. How general the mechanisms we identify here are to other instances of dissent is, however, a question that will require further research.

Figure1 . Geographical distribution of the protests and intensity of online activity

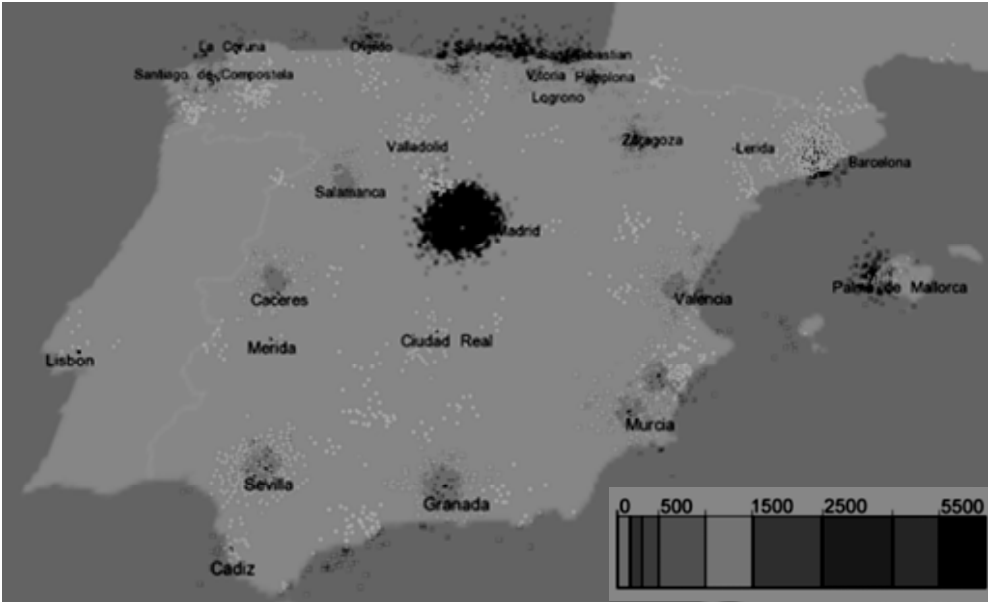



Figure 2. Most popular #hashtags before and after the first demonstration day (in bold)

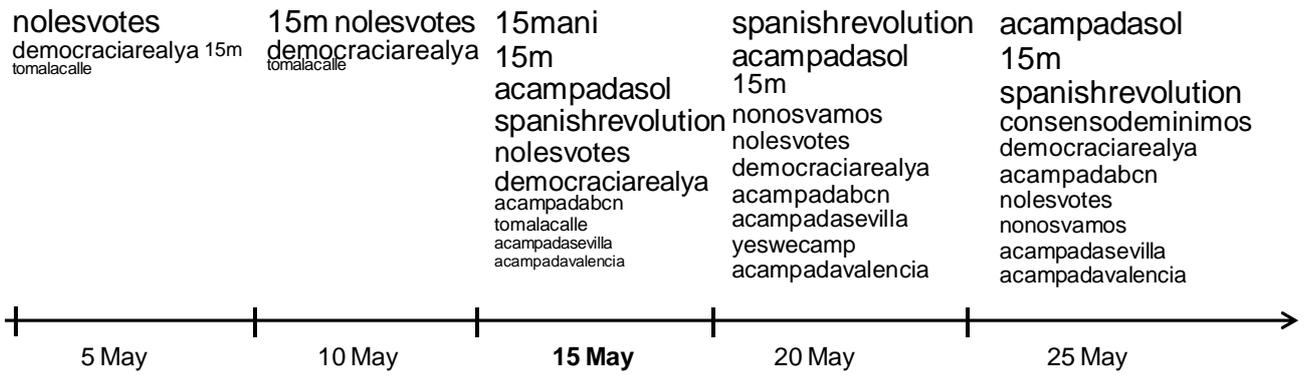



Figure 3. Online growth of the movement and offline media coverage

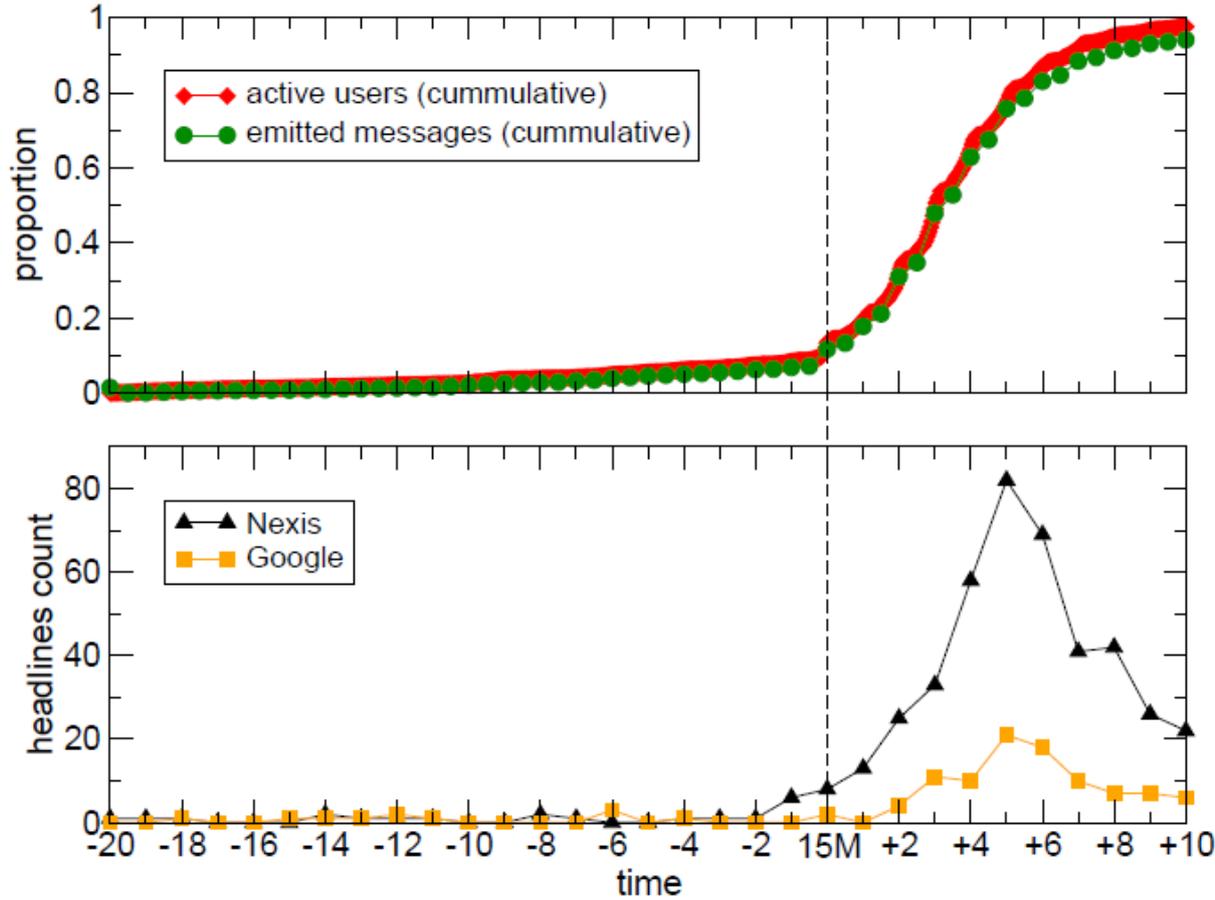



Figure 4.  Distribution of users according to network position and message activity

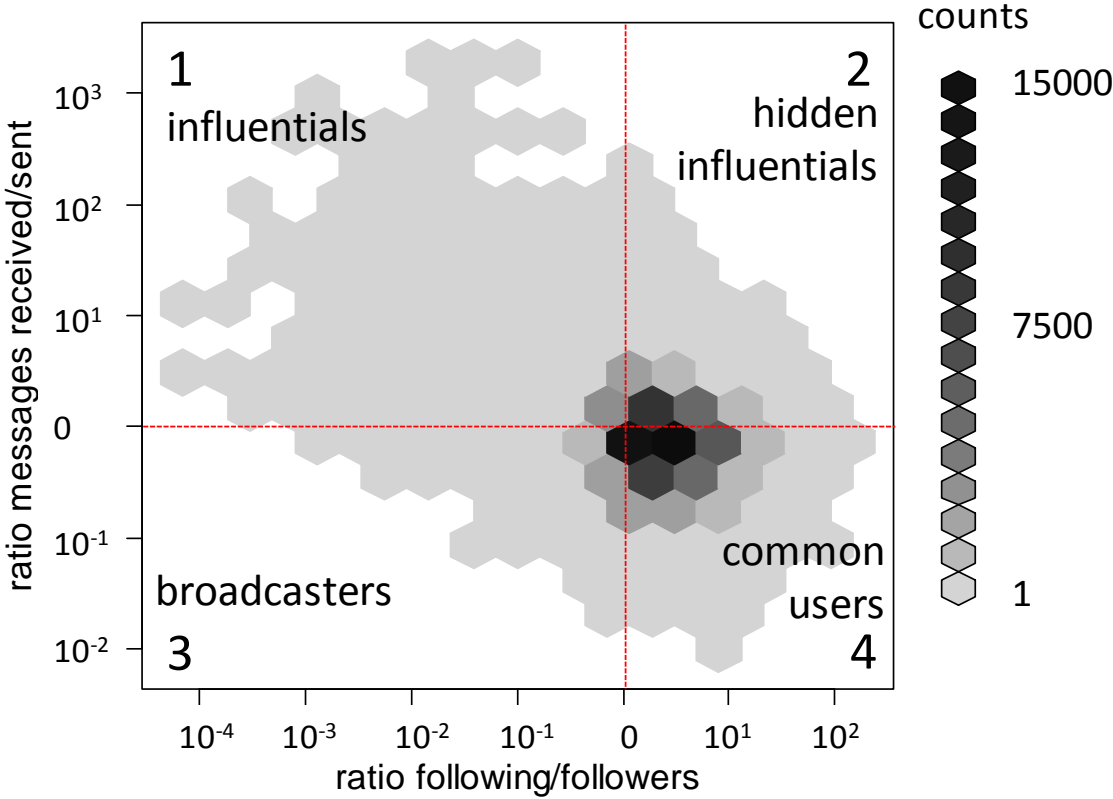



Figure 5. Distribution of activation time and cascade sizes

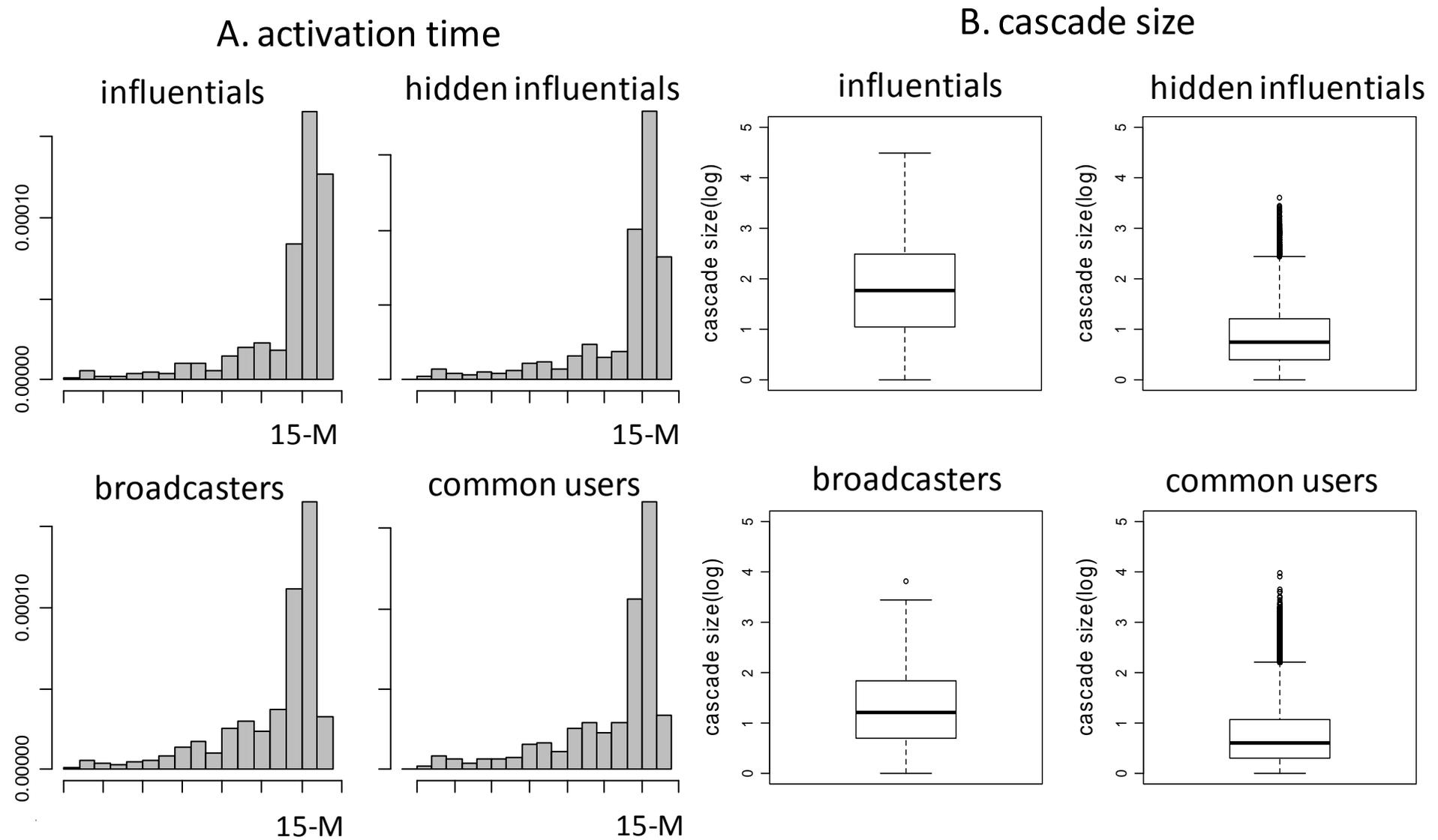



Table 1. Ranking of top 20 users in terms of network centrality and message visibility

| | global top 20 | | top 20 message visibility | | | |
|---|---|---|---|---|---|---|
| | network centrality | message visibility | influentials | hidden influentials | broadcasters | common users |
| 1 | *celebrity* | *political party* | *political party* | **Democra_Real_Ya** | *personal account* | *personal account* |
| 2 | wikileaks | twitpic *(app)* | twitpic *(app)* | **nonosvamos** | *personal account* | *personal account* |
| 3 | *blog* | *national newspaper* | *national newspaper* | **indignaos** | *personal account* | *personal account* |
| 4 | twitter_es | youtube | youtube | **15m** | *personal account* | *personal account* |
| 5 | reuters | *national news channel* | *national news channel* | *personal account* | vudeo_org | *personal account* |
| 6 | *police* | *news website* | *news website* | *personal account* | *personal account* | *personal account* |
| 7 | *celebrity* | *celebrity* | *celebrity* | ikiMap *(app)* | *personal account* | *personal account* |
| 8 | Twibbon *(app)* | *celebrity* | *celebrity* | *personal account* | *celebrity* | **Acampadalisboa** |
| 9 | *national newspaper* | *celebrity* | *celebrity* | *regional newspaper* | *personal account* | *personal account* |
| 10 | twitter support | AddThis *(app)* | AddThis *(app)* | *local politician* | *celebrity* | *personal account* |
| 11 | *celebrity* | **acampadasol** | **acampadasol** | *personal account* | *personal account* | *personal account* |
| 12 | *procrastination* | *news website* | *news website* | **acampadazamora** | EPMadrid | *personal account* |
| 13 | NYtimes | europapress_es | europapress_es | *personal account* | *personal account* | *personal account* |
| 14 | *news website* | *regional politician* | *regional politician* | *personal account* | *personal account* | *personal account* |
| 15 | cnnbrk | *celebrity* | *celebrity* | *personal account* | *personal account* | *personal account* |
| 16 | yfrog *(app)* | *national newspaper* | *national newspaper* | *local political party* | *personal account* | *personal account* |
| 17 | ***Nolesvotes*** | *national newspaper* | *national newspaper* | *personal account* | *celebrity* | *personal account* |
| 18 | FCBarcelona_es | *celebrity* | *celebrity* | Swivel *(app)* | *personal account* | *personal account* |
| 19 | *procrastination* | *celebrity* | *celebrity* | **SpainRevolt** | *personal account* | *personal account* |
| 20 | *international politician* | *celebrity* | *celebrity* | *personal account* | *personal account* | *personal account* |

Note: labels in italics substitute the original usernames to preserve anonymity or assign general categories to actors or organizations not known internationally. Usernames in bold were created as part of the protest campaign. The abbreviation *app* in parenthesis identifies Twitter applications.